\documentclass[prB,amsmath,amssymb,twocolumn]{revtex4}
\usepackage[normalem]{ulem}
\usepackage{amsmath,amssymb}
\usepackage[ps2pdf,bookmarks=true,colorlinks,linkcolor=red,urlcolor=blue,citecolor=blue]{hyperref}
\usepackage{graphicx,amsmath,amsfonts}
\usepackage[usenames]{color}
\usepackage{epstopdf}

\newcommand{\be}{\begin{equation}}
\newcommand{\beq}{\begin{eqnarray}}
\newcommand{\eeq}{\end{eqnarray}}

\def \be{\begin{equation}}
\def \ee{\end{equation}}
\def \ba{\begin{array}}
\def \ea{\end{array}}
\def \bea{\begin{eqnarray}}
\def \eea{\end{eqnarray}}
\def \nn{\nonumber}
\def \half{\frac{1}{2}}

\def \bq{{\bf q}}

\def \e{{\epsilon}}

\def \a{{\alpha}}
\def \t{{\theta}}

\def \d{{\delta}}
\def \w{{\omega}}

\def \x{{\chi}}
\def \e{{\epsilon}}

\def \av#1{{\langle#1\rangle}}

\begin{document}

\title{Quantum critical states and phase transitions in the presence of non equilibrium noise}
\author{Emanuele G. Dalla Torre$^1$, Eugene Demler$^2$, Thierry Giamarchi$^3$, Ehud Altman$^1$}
\affiliation{
$^1$ Department of Condensed Matter Physics, Weizmann Institute of Science,
Rehovot, 76100, Israel\\
$^2$Department of Physics, Harvard University, Cambridge MA
02138\\
$^3$DPMC-MaNEP, University of Geneva, 24 Quai Ernest-Ansermet, 1211 Geneva,
Switzerland}


\begin{abstract}
Quantum critical points are characterized by scale invariant correlations and correspondingly long ranged entanglement.
As such, they present fascinating examples of quantum states of matter, the study of which has been an important theme in modern physics.
Nevertheless very little is known about the fate of quantum criticality under non equilibrium conditions. In this paper we investigate the effect of
external noise sources on quantum critical points. It is natural to expect that noise will have a similar effect to finite temperature, destroying the subtle correlations
underlying the quantum critical behavior. Surprisingly we find that the ubiquitous $1/f$ noise does preserve the critical correlations. The emergent states show intriguing interplay of intrinsic quantum critical and external noise driven fluctuations. We demonstrate this general phenomenon with specific examples in solid state and ultracold atomic systems. Moreover our approach shows that genuine quantum phase transitions can exist even under non equilibrium conditions.
\end{abstract}

\maketitle
\section{Introduction}
An important motivation for investigating the behavior of non-equilibrium quantum states comes from state of the art experiments in atomic physics. Of particular interest in this regard are
systems of ultracold polar molecules\cite{Polar1,Polar2} and long chains of ultracold trapped ions\cite{IonTraps}. On the one hand these systems offer unique possibilities to realize strongly correlated many-body states, which undergo interesting quantum phase transitions\cite{PolarReview,Porras,Mata}. But on the other hand they are controlled by large external
electric fields, which are inherently noisy and easily drive the system out of equilibrium\cite{MonroeNoise,ChuangNoise}. It is natural to ask what remains of the quantum states, and in particular, the critical behavior under such conditions.

The effect of non-equilibrium noise on quantum critical points is also relevant to more traditional solid-state systems. Josephson junctions, for example, are known to be affected by non-equilibrium circuit noise, such as $1/f$ noise. Without this noise a single quantum Josephson junction should undergo a text-book quantum phase transition\cite{Schmidt,Chakravarti}: depending on the value of a shunt resistor, the junction can be in either a normal or a superconducting state. A phase transition occurs at a universal value of the shunt resistance $R_s=R_Q=h/(2e)^2$, independent of the strength of the Josephson coupling. This is closely related to the problem of macroscopic quantum tunneling of a two level system (or q-bit) coupled to a dissipative environment\cite{LeggettReview}.

There is a large body of work on 1/f noise as a source of decoherence for superconducting q-bits (see e.g. \cite{ClarkReview,Schoen}).
However the effect of such noise on the quantum phase transitions and the non equilibrium steady states of Josephson junctions poses fundamental open questions. Do the different phases (superconducting or normal) retain their integrity  in presence of the noise? Is the phase transition between them sharply defined?

In certain cases it was argued that a non equilibrium drive
may act as an effective temperature\cite{Mitra, DiehlZoller}.
And temperature is known to be a relevant perturbation, which
destroys quantum criticality\cite{SachdevBook,SondhiReview}.
In contrast, we find that the external $1/f$ noise is only a {\em marginal} perturbation at the critical point in many cases of interest. This is exemplified in Sec. \ref{sec:JJ} for a shunted Josephson junction subject to charge noise. In section \ref{sec:1D} we investigate the potentially richer physics of one-dimensional systems of
ultra cold polar molecules or trapped ions. These systems form a critical state at $T=0$,
which can undergo pinning in the presence of a commensurate lattice or a single impurity.
Pinning occurs as a quantum phase transition at a critical value of the correlation
exponent\cite{giamarchi} (For application to ion traps see Ref. [\onlinecite{Mata}]).
Another interesting phenomenon in ion chains is the zigzag instability\cite{MorigiFishman}, which is expected to evolve into a true quantum phase transition in the limit of long chains.

Again the relevant issue
is the fate of these critical states and quantum phase transitions in the presence of noisy electrodes.
Such noise has been characterized in recent experiments with ion traps\cite{MonroeNoise,ChuangNoise}, where it was found to have a $1/f$ power spectrum and attributed to localized charge patches on the electrodes.
A crucial result of our analysis is that such noise preserves the critical states,
and the exponents are continuously tuned by it.
The fact that the system is out of equilibrium is betrayed by the linear response to an external probe, such as light scattering. The energy dissipation function of the scattered light can become negative for sufficiently strong external noise, exhibiting gain instead of loss.

The long wavelength description of the noise-driven steady state allows us to study its stability
to various static perturbations within a renormalization group framework. In this way we describe pinning by a static
impurity and by a lattice potential. We show that pinning-depinning occurs as a phase transition driven by interplay of the intrinsic quantum fluctuations and the external noise. Before proceeding we note previous work which found
modified quantum criticality in cases where the non equilibrium conditions were due to an imposed current\cite{Racz,Feldman}.

\begin{figure*}[t]
\includegraphics[scale=1.0]{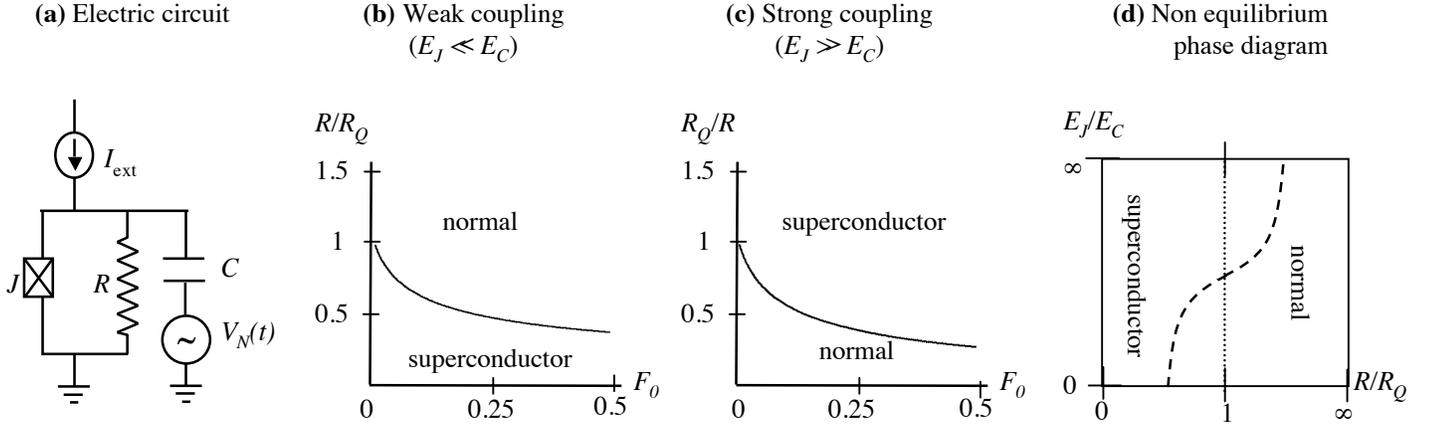}
\caption{Effects of non-equilibrium noise on the localization quantum phase transition of a single shunted Josephson junction: {\bf(a)} Electronic circuit relevant to a resistively shunted Josephson junction with charging noise. {\bf(b)} Critical resistance $R/R_Q$ as function of the noise strength $F_0$, in the weak coupling limit. {\bf (c)} Critical conductance $R/R_Q$ as function of the noise strength $F_0$, in the strong coupling limit. Figures (a) and (b) are related by the duality transformation $R/R_Q\to R_Q/R$ and "superconductor"$\leftrightarrow$"insulator". {\bf (d)} Schematic phase diagram at equilibrium (dotted line) and in the presence of non-equilibrium one-over-f noise (dashed line)}
\label{fig:phdia4}
\end{figure*}

\section{Phase transition in a noisy Josephson junction}\label{sec:JJ}

In our discussion of the Josephson junction we consider the standard circuit shown in Fig. \ref{fig:phdia4}(a).
The offset charge $eN_0$ on the capacitor has random time dependent fluctuations with a $1/f$ spectrum \cite{1overfnoiseJJ} $\av{N_0^*(\w)N_0(\w)} = F_0/|\w|$. This is modeled by the fluctuating voltage source $V_N(t)=eN_0(t)/C$.


{\em Weak coupling --} Consider first the system
at vanishing Josephson coupling, which is then just an $RC$ circuit. Treating the resistor as
an ohmic bath in thermal contact with the system\cite{CalderiaLeggett}
results in the Langevin equation for a damped quantum oscillator:
\be
\half c \ddot\t+ \eta \dot\t=\zeta(t)+ \half\dot N_0(t).
\label{langevinJJ}
\ee
Here $c=\hbar C/2e^2$ and $\eta=(1/2\pi) R_Q/R$.
The random forcing term
$\zeta(t)$ originates from the equilibrium bath, and therefore at $T=0$ has
the power spectrum $\av{\zeta^\star_\w\zeta_\w}=\eta |\w|$. The other random forcing term is the time derivative of the charge noise. Since the charge fluctuations have a
spectrum $\sim F_0/|\w|$, the power spectrum of $\dot N_0$ is $\sim F_0|\w|$, which mimics the resistor noise. Unlike the resistor, however, external fluctuations do not have an associated dissipation term. This is because the noise source is not in thermal contact with the system. Thus the fluctuation dissipation theorem is violated in the presence of the non-equilibrium noise source.

Using the linear equation of motion (\ref{langevinJJ}) we can compute the phase autocorrelation function:
\bea
\av{\cos\left[\t(t)-\t(0)\right]}\sim t^{-(1+F_0/\eta)/\pi\eta}
\label{autocorr}
\eea
Interestingly, the non equilibrium noise does not destroy the power-law scaling, but modifies the exponent. We conclude that a critical (scale-invariant) non-equilibrium steady state obtains in presence of the external noise.

The important question to address in the context of weak coupling, is under what conditions the critical steady-state we just described is stable to introduction of the Josephson coupling as a perturbation. That
is, we should find how the perturbation transforms under
a scale transformation that leaves the critical steady state
invariant.
To investigate this we turn to formulation of the dynamics
in terms of the Keldysh action described in the methods section. The
quadratic action (\ref{S0JJ}) describing the $RC$ circuit is scale invariant, whereas
the Josephson coupling term
\be
S_J=J\int dt\left[\cos\t_f(t)-\cos\t_b(t)\right]
\label{SJ}
\ee
is not in general. Here $\t_f$ ($\t_b$) is the field on the forward (backward) part of the Keldysh contour.
From the decay of the correlation function
(\ref{autocorr}) we can directly read off the anomalous scaling dimension of the perturbation, which is
$\a=1-(1+ F_0/\eta)/2\pi\eta$. When $\a>0$ the perturbation grows under renormalization
and ultimately destabilizes the critical steady-state.
We therefore predict a phase transition
at a critical resistance ${R^*/R_Q}= \left(\sqrt{8\pi F_0+1}-1\right)/4\pi F_0$,
below which, the Josephson coupling term becomes relevant.
Note that we recover the equilibrium dissipative transition at $R^*=R_Q$
in a "quiet" circuit ($F_0=0$). We can tune across the transition
also by maintaining a constant resistance $R<R_Q$ and increasing
the non-equilibrium  noise "power" $F_0$, as shown in Figure \ref{fig:phdia4} (b).

Within the weak coupling theory we do not have direct access to the properties
of the steady-state at $R<R^*$. However because the Josephson coupling grows under renormalization it is reasonable to expect that the junction
would be superconducting. To determine this with more confidence we shall now take the opposite, strong coupling
viewpoint.

{\em Strong coupling --}  We employ a well known duality between weak and strong coupling
\cite{Schmidt,FisherZwerger}, under which
cooper pair tunneling $J\int  dt~\cos(\t)$ is mapped to tunneling of phase slips across the
junction $S_g = g\int dt~\cos(\phi)$. Concomitantly the resistance $R/R_Q$ is mapped to
a normalized conductance $R_Q/R$.  In the strong coupling limit of the Josephson junction $J>>e^2/c$, the dual action,
with a phase slip tunneling $S_g$, is at weak coupling.
The scaling analysis can proceed in the same way as above, giving a transition at the value of shunt resistance $R^*/R_Q = 4\pi F_0/(\sqrt{8\pi F_0+1}-1)$. For $R<R^*$ the phase-slip tunneling $S_g$ is irrelevant. That is at asymptotically long times all phase slip events
making the superconducting state stable for $R<R^*$, at least in the strong coupling limit.

The combined results of the weak and strong coupling analysis imply a phase diagram of the form shown in
Figure \ref{fig:phdia4}(d). At weak coupling the critical resistance, in presence of noise, occurs at $R^*$
which is smaller than $R_Q$, while at strong coupling $R^*$ is larger than $R_Q$. The dashed line in this figure shows a simple interpolation of the phase boundary between the two limiting regimes. However, we cannot exclude the possibility that new phases, such as a metallic phase arise at intermediate coupling.

\section{One dimensional chains of polar molecules or trapped ions}\label{sec:1D}

We now turn to investigate the interplay between critical quantum fluctuations and external classical noise in one-dimensional systems. Good laboratories for studying such effects are ions in Ring or linear Paul traps, as well as Polar molecules confined to one dimension. Because of the confinement to one dimension both systems are affected by quantum fluctuations. On the other hand they are also subject to noisy electric fields that can influence the steady state correlations.

In ion traps, the fluctuations of the electric potential, which couples to the ionic charge density have been carefully characterized\cite{MonroeNoise,ChuangNoise}.
The noise power spectrum was found to be very close to $1/f$ and with spatial structure indicating moderately short range correlations. In the molecule system electric fields are used to polarize the molecules, and fluctuations in these fields couple to the molecule density via the molecular polarizability.

Our starting point for theoretical analysis is the universal harmonic theory describing
long wavelength phonons in the one dimensional system\cite{Haldane}, which is written in terms of the
displacement field $\phi(x,t)$ of the particles from a putative Wigner lattice.
The long wavelength density fluctuations are represented by
the gradient of the displacement field,
$(-1/\pi)\partial_x\phi(x,t)$. The part of the density with fourier components of wavelengths near the inter-particle spacing are encoded by  ${\hat O}_{DW}= \rho_0\cos(2\pi\rho_0 x + 2\phi(x,t))$\cite{Haldane,giamarchi}, where $\rho_0$ is the average density. The operator $O_{\rm DW} = \rho_0\cos(2\phi(x,t))$ is density wave (or solid) order parameter field of the Wigner lattice.
As in the Josephson junction, we wish to address two questions; (i) How does the external noise affect the steady-state, which in equilibrium exhibits algebraic correlations; (ii) How does it influence phase transitions, such as the lattice pinning transition.

We model the external electric noise
as a random time dependant field coupled to the particle density. In general, the noise couples to both components of the density via the terms
$-f(x,t)\pi^{-1}\partial_x\phi(x,t)$ and $\zeta(x,t)\rho_0\cos(2\phi(x,t))$. For now we assume that the noise source is correlated over sufficiently long distances, that its component at spatial frequencies near the particle density ($\zeta(x,t)$) is very small and can be neglected. In this case the long wave-length theory remains harmonic. We shall characterize the noise by its power spectrum $F(q,\w)=\av{f(q,\w)f(-q,-\w)}$. We take this to be $1/f$ noise with short range spatial correlations,
that is $F(q,\w)= F_0/|\w|$.

When the system is irradiated with external noise we expect it to absorb energy. In order to stabilize a steady state we need a dissipative bath that can take this energy from the system. In the Josephson junction problem, the resistor naturally played this role. Is there a similar dissipative coupling in the one dimensional systems under consideration here?

In the ion traps, there is a natural dissipative coupling because these systems can be continuously laser cooled.  Thus the system can reach a steady state, which reflects a balance between the laser cooling and the external noise (see Appendix A). The polar molecules do not couple to a natural source of dissipation, however a thermal bath can in principle be realized by immersion in a large atomic condensate\cite{Zoller_immerse}. In the Appendix we show that the bath generated by a two dimensional weakly interacting condensate provides
the needed dissipation.

The combined effects of interactions, external noise and dissipation, are described by a quadratic Keldysh action as shown in the methods section (\ref{S_Keldysh}). This is the natural extension from the single junction (\ref{S0JJ})
to the one dimensional chain. But there is an important difference. The harmonic chain, is scale invariant only without the noise and
dissipation terms, which are strictly speaking relevant perturbations of this fixed point.
Indeed, the dissipative coupling generates a relaxation time-scale $\tau\sim 1/\eta$,
which breaks the scale invariance. To retain the scale invariance and still drive the system out of equilibrium we can consider the interesting limiting regime in which both $\eta\to 0$ and $F_0\to 0$, while the ratio $F_0/\eta$ tends to a constant. Then the correlation function is easily calculated and seen to be a power law
\be
\av{\cos(2\phi_{cl}(x))\cos(2\phi_{cl}(0))}\sim x^{-2K(1+\pi^{-2} F_0/\eta)}
\label{Crystal}
\ee
where $K$ is the Luttinger parameter, which determines the decay of correlations at equilibrium ($F_0=0$).
The same exponent holds for the temporal correlations. We see that the dimensionless ratio
$F_0/\eta$, which measures the deviation from equilibrium, acts as a marginal perturbation.
In practice $\eta$ and $F_0$ are non vanishing. Then the result (\ref{Crystal}) will be valid at scales shorter than $1/\eta$. Correlations will decay exponentially at longer scales.
Thus $\eta$ serves as an infrared cutoff of the critical steady-state. In practice however
the system size or cutoff of the $1/f$ spectrum may set more stringent infrared cutoffs.

The density-density correlations can be measured directly by light scattering. The (energy integrated) light diffraction pattern in the far field limit gives directly the static structure factor $S(q)=\av{\rho_{-\bq}\rho_\bq}$ of the sample. In particular, the power-law singularity in $S(q)$ near wave-vector $q_0\sim 2\pi\rho_0$ is just the Fourier transform of the power law decay of the Wigner crystal correlations (\ref{Crystal}).

We can also compute the decay of phase correlations $\av{{\cos[\t(x)-\t(0)]}}$, which in the system
of cold molecules may be measured by interference experiments [\onlinecite{AnatoliPNAS}].
By considering the dual representation of the harmonic action (\ref{S_Keldysh}) we find a decay
exponent $(1+ F_0/\eta)/2K$.

At equilibrium both crystalline and phase correlations are controlled by $K$ alone:
reducing $K$ (by increasing interactions) leads to a slower decay of density-wave correlations and concomitantly faster decay of phase correlations. This duality, a consequence of minimal uncertainty between
phase and density in the harmonic ground state, is violated in the presence of noise.
Increasing the noise leads to a faster decay of both the density and phase correlations.


\subsection{Response}\label{sec:response}


\begin{figure}[t]
\includegraphics[width=8cm]{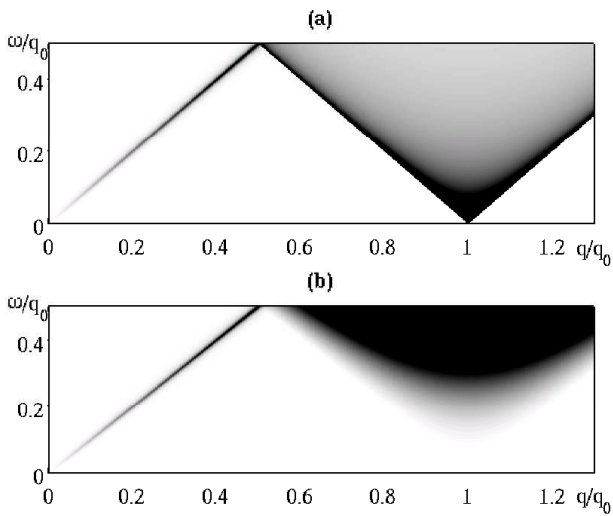}
\caption{Effects of non-equilibrium noise on the response to Bragg spectroscopy: {\bf (a)} Imaginary part of the response function $\chi''(q,\w)$ in a one dimensional system with $K=0.5$, at equilibrium ($F_0=0$). {\bf (b)} Same plot, in the presence of a strong $1/f$ noise with $F_0/\eta=4\pi^2$.}
\label{fig:response}
\end{figure}

Under the non-equilibrium conditions, the fluctuation dissipation theorem does not hold in general and we should consider response functions separately from the correlations. Here we discuss the density-density response, whose fourier-transform gives the linear response of the density of the system to a dynamic perturbation consisting of a weak periodic potential with wave-vector $q$, oscillating at a frequency $\w$. This is the response function probed by Bragg spectroscopy\cite{BraggKetterle,BraggNir,BraggInguscio}.

The combined response at small wave-vectors $q\ll q_0=2\pi\rho_0$  is unmodified by the noise.
This is because the probe field at couples linearly to $\phi$ via the smooth part of the density $\sim \partial_x\phi$. Since the system is harmonic, any two perturbations that couple linearly to the oscillator field simply add up independently. Hence we have $\chi''(q,\w)=K|q|\Theta(\w)\d(\w-q)$  as in equilibrium.
On the other hand, the response at wave-vectors near the inverse inter-particle distance involves a non-linear coupling via the  component of the density $\rho_{q_0}(x,t)=\cos(q_0x+2\phi)$.
We calculate this exactly and obtain (see Appendix B)
\bea
\chi''(q,\w)&=&C\left(K,K_\star\right)(\w^2-\d q^2)^{K_\star-1}\Theta(\w^2-\d q^2)\nn\\
C(K,K_\star)&=&{1\over 4\Gamma^2(K_\star)}{\sin(\pi K)\over\sin(\pi K_\star)}
\label{chipp}
\eea
Here $\Gamma$ is the Gamma function, $\d q\equiv q-q_0$, and we have defined $K_\star\equiv K(1+\pi^{-2} F_0/\eta)$.

The response near $q=0$ and $q=q_0$ is shown in
Fig. \ref{fig:response}.
By comparing the plot in panel (a), showing the case of vanishing noise, to panel (b) where $F_0/\eta=4\pi^2$,
we see that the noise turned the divergent response at $q=q_0$ to a power law suppression.
What betrays the fact that the spectrum (b) stems from non-equilibrium conditions and not just
weaker interactions? Consider the relation between the the response and the energy dissipation
functions $\dot E(q,\w)=\w \chi''(q,\w)$, i.e. the rate by which the probe is doing work on the system.
Inspecting the pre-factor $C(K,K_\star)$ in Eq. (\ref{chipp}) we find, that for sufficiently strong noise $F_0/\eta> \pi^2(1-K)/K$ (for $K<1$), the
energy dissipated by the probe can become negative, which would be strictly prohibited if the system was at equilibrium.

The situation is analogous to a laser, where gain is achieved by pumping the medium out of equilibrium to ``population inversion''. Here the external $1/f$ noise
plays the role of the pump. In contrast to a laser the gain spectrum is continuous and reflects the critical properties of the many-body steady state. Eq. (\ref{chipp}) implies a commensurability effect between the noise and the intrinsic interactions that leads to oscillations between gain and loss as a function of the noise power.

\subsection{Non equilibrium phase transitions}\label{sec:SF-MI}

We have seen that by changing the $1/f$ noise one can continuously tune the critical exponent associated
with the power-law decay of correlations in one dimensional quantum systems.
As in the case of the Josephson junction,
we can ask if it is possible to use the new knob to tune across a phase transition.

A text book\cite{giamarchi} phase transition in one dimensional quantum systems is that of pinning by a commensurate periodic lattice potential. In equilibrium it occurs below a universal critical value of the Luttinger parameter $K_c=2$, regardless of the strength of the potential. In the context of the real time dynamics,
the periodic potential is added as a perturbation to the action (\ref{S_Keldysh})
\be
S_g=g\int dx dt ~\cos(2\phi(x,t))
\label{Sp}
\ee
The scaling of the perturbation (\ref{Sp}) in the critical steady state
is determined with the help of the
correlation function (\ref{Crystal}). We find that the action of the periodic lattice has
the scaling dimension $\a_p=2-K(1+\pi^{-2}F_0/\eta)$.
This implies an instability, which signals a phase transition to a pinned state
for $ F_0/\eta < \pi^2(2K^{-1}-1)$. In particular
for $F_0=0$ we recover the equilibrium pinning (or Mott)
transition at the universal value of the Luttinger parameter $K_c=2$. Note that for $K>2$ the system is always unpinned because $F_0$ is non-negative.

A pinning transition can also occur in the presence of a single impurity \cite{KaneFisher}.
The main difference from the previous case is that this perturbation is completely local and therefore its
scaling dimension is reduced by $1$ relative to the periodic potential: $\a_i=1-K(1+\pi^{-2} F_0/\eta)$.
Accordingly the de-pinning transition occurs at a lower critical noise
$ F_0/\eta=\pi^2(K^{-1}-1)$, than in the case of the periodic potential.


\section{Discussion and Conclusions}

We described a new class of non-equilibrium quantum critical states and phase transitions, which emerge in
the presence of external classical noise sources. Physical examples include a Josephson junction
and one dimensional chains of trapped ions or polar molecules coupled to $1/f$ noise.
In contrast to thermal noise, which destroys quantum criticality, the $1/f$ noise preserves the
algebraic decay of correlations and thus acts as a {\em marginal}
perturbation at the quantum critical point in these systems.
A noise that deviates from $1/f$ at low frequencies, e.g. $1/f^{1+\e}$, is relevant (irrelevant) for $\e>0$ ($\e<0$).
However, for $|\e|<<1$ the critical correlations will be maintained below the crossover scale $t_*\sim t_0 \exp(1/|\e|)$, where $t_0$ is the short time cutoff.

The critical exponents associated with both phase and density correlations are varied continuously by the noise,
which also destroys the well known duality between the two. An even more dramatic effect of the
non-equilibrium conditions  is betrayed by the dissipative response of the critical steady-state
to an external probe field, which for strong noise can change sign and turn from loss to gain.

Quantum phase transitions, such as pinning of the crystal
by an impurity or by a commensurate lattice potential can take place in the presence of the external, non-equilibrium
noise. In particular the system can be tuned across the depinning transition by tuning the noise power.

It would be interesting to extend these ideas to higher dimensional systems, such as
one or two dimensional arrays of coupled tubes of polar molecules.
The natural phases in equilibrium are the broken symmetry phases, either
superfluid or charge density wave. The intriguing sliding Luttinger liquid phase, which
retains the one dimensional power-law correlations despite the higher dimensional coupling,
is expected to be stable only in a narrow parameter regime\cite{KollathMeyer}.
Because the $1/f$ noise acts to suppress both the phase and density correlations it
will act to stabilize this phase in a much wider regime. It would also be interesting
to consider the effect of the noise on more complex phase transitions,
such as the zigzag instability of ion chains\cite{MorigiFishman} as well as Josephson junction arrays\cite{GilJJA}.

{\em Acknowledgements.} We thank Erez Berg, Sebastian Huber, Steve Kivelson, Austen Lamacraft, Kathryn Moler and Eli Zeldov for stimulating discussions.
This work was partially supported by the US-Israel BSF (EA and ED), ISF (EA), SNF under MaNEP and division II (TG), E. D. acknowledges support from NSF DMR-0705472, CUA, DARPA, and MURI. E.G.D.T. is supported by the Adams Fellowship Program of the Israel Academy of Sciences and Humanities.

\section{Methods}

\paragraph{Keldysh action of the quantum Josephson Junction}
The linear quantum Langevin equation (\ref{langevinJJ}) is equivalent to the quadratic Keldysh action \cite{Kamenev}
\be
S_0 =\sum_{\w,q} \ba{c c}(\t_{cl}^* & {\hat \t}^*)\\ & \ea \left(\ba{c c} 0 & \half c \w^2 -i \eta\w \\ \half c \w^2+i \eta\w & -2i \eta|\w| -2i\w^2\frac{ F_0}{|\w|} \ea \right)\left(\ba{c}\t_{cl}\\ {\hat \t}\ea\right)
\label{S0JJ}
\ee
Here $\t_{cl}$, and $\hat\t$ are the "classical" and "quantum" fields. As usual they are defined as
the symmetric and anti-symmetric combinations, respectively, of the fields associated with forward and backward time propagation of operators: $\t_{cl}=(\t_f+\t_b)/2$, ${\hat\t}=\t_f-\t_b$. The josephson coupling (\ref{SJ}) is added to this action.

We note that the contribution of the non equilibrium noise has the same scaling dimension as the terms coming from the resistor $\propto |\w|$.
By contrast the capacitive term $\propto\w^2$ is irrelevant, at low frequencies. As a result, the
fixed-point action, governing the exponent of (\ref{autocorr}), does not depend on $c$.

\paragraph{Keldysh action for one-dimensional systems}
The Keldysh action that describes the long wavelength density fluctuations, coupled to
the external noise and the dissipative bath is given by
\begin{widetext}
\bea
S_0 =\sum_{\w,q} \ba{c c}(\phi_{cl}^* & {\hat \phi}^*)\\ & \ea \left(\ba{c c} 0 & {1\over\pi K}(\w^2-q^2)-i \eta\w \\ {1\over\pi K}(\w^2-q^2)+i \eta\w & -2i \eta|\w| -2i \frac{q^2}{\pi^2}\frac{ F_0}{ |\w|} \ea \right)\left(\ba{c}\phi_{cl}\\ {\hat \phi}\ea\right)
\label{S_Keldysh}
\eea
\end{widetext}
Here $F_0/|\w|$ is power spectrum of the external noise. The factor of $q^2/\pi^2$ in front of this term appears  because the noise couples to $(1/\pi)\partial_x\phi$, the smooth part of the density. $\eta$ denotes the dissipative coupling, which is derived in Appendix A.

\appendix
\section{Microscopic model for the dissipative bath}

To reach a steady state, the models studied in the paper require coupling to a dissipative bath, that can take away the excess energy produced by the external noise sources. In the Josephson junction the
dissipation was naturally provided by a resistor. In the one dimensional systems which we addressed,
the dissipation must be provided by some kind of cooling apparatus. In the ion chain this is done with continuous laser cooling. For the system of dipolar molecules
we proposed to realize the required dissipation by sympathetic cooling, via immersion in a large condensate.

In this appendix we describe the two methods in some detail and show that they indeed lead to the required dissipative term in the action, proportional to $|\w|$.

\subsection*{Laser Cooling}

The motion of a laser cooled ion is described\cite{CT_laser} by the Langevin equation \be \dot{p}_i(t)=-\gamma p_i(t)+F_i(t)+\zeta_i(t)\label{Langevin},\ee
where $p_i(t)$ is the momenutum of the ion, $\gamma$ the damping rate (in the case of doppler cooling $\gamma$ equals half the recoil energy). $F_i(t)$ is the force affected on the ion $i$ by the confining potential and the neighboring ions. For example the harmonic force from the neighboring ions would be $K(x_{i+1}+x_{i-1}-2x_i)$. The stochastic force $\zeta(t)$ stems from spontaneous emission events. Note that both the dissipation $\gamma$ and the fluctuations $\zeta_i(t)$ are completely local. This is because they originate from absorption and emission of photons of very short wavelength (much shorter than the inter ion distance) corresponding to an atomic transition. Moreover, the master equation associated with laser cooling alone (i.e. in absence of any external noise) has only one attractive steady state solution, which is a thermal state of the harmonic oscillators at some effective temperature $T_{eff}$ \cite{CT_laser}. The quantum recoil noise due to spontaneous emissions must therefore
have the thermal spectrum $\av{\zeta^\star_{\w}\zeta_{\w}} = 2\gamma ~T_{eff}\coth(\w/ 2T_eff)$.

The Langevin equations are easily transferred to the continuum limit of the ion chain. The momentum $p(t)$ becomes the field $\Pi(x,t)$, canonically conjugate to the ion
field $\phi(x,t)$, which represents the displacements of the ions from their putative lattice. Accordingly, the harmonic interaction between ions in the chain leads to the force $F \to -(1/\pi K)\partial_x^2\phi(x,t)$.
Next we recast the harmonic Langevin equations in the form of a Keldysh action
in the standard way\cite{Kamenev}. For ground state cooling, i.e. in the limit $T_{eff}<<\w_{min}$ the action can be written as
\begin{widetext}\be
S_0 =\sum_{\w,q} \ba{c c}(\phi^* & {\hat \phi}^*)\\ & \ea \left(\ba{c c} 0 & {1\over\pi K}(\w^2-q^2)-i \gamma\w \\ {1\over\pi K}(\w^2-q^2)+i \gamma\w & -2i \gamma|\w| \ea \right)\left(\ba{c}\phi\\ {\hat \phi}\ea\right).
\label{S_Keldysh2}\ee\end{widetext}
Of course, to this action we must independently add the external  $1/f$ noise
which will lead to the action of Eq. (8) in the methods section.
We see that the establishment of laser cooling as a thermal dissipative bath stems directly from (i) the locality of the cooling process, and (ii) the fact that
Laser cooling acting alone (without the external noise) leads to a thermal state.

\subsection*{Immersion in a condensate}


A practical way to provide a dissipation (and cooling) mechanism in a one dimensional system of ultracold molecules is to immerse it in a large atomic condensate.
The BEC is in turn evaporatively cooled in the standard way. We will show below, that within a wide regime this scheme indeed leads to the local ohmic dissipation (i.e. $\propto |\w|$ and independent of wave-vector), needed to balance the $1/f$ noise. The crucial conditions which need to be satisfied to obtain local ohmic dissipation
as required,
are (i) that the atomic condensate is two dimensional; (ii) that its sound velocity is much smaller than the velocity of phonons in the molecular chain.

The coupling between the atoms and the molecules is via a density-density interaction
$H_{am} =g\int d^d r \d\rho_m(r) \d\rho_a(r)$. To derive the dissipative term for
the molecules we integrate out the density fluctuations of the
atomic condensate. This results in a coupling term in the effective
action for the molecules
\be
S_{diss}={g^2\over 2}\int dr dr' \d\rho_m(r')\x_a(r'-r)\d\rho_m(r).
\label{Sm}
\ee
Here and in what follows we employ the imaginary time (Matsubara) formalism and
we have denoted $r\equiv(x,\tau)$. The interaction is mediated by $\chi_a(r)$, the density density correlation function, or phonon propagator,
in the two dimensional atomic condensate. In Fourier space: $\tilde\chi_a= q^2/(\w^2+v_a^2 q^2)$, where $v_a$ is the sound velocity in the atomic condensate.

To make further progress we turn to the long-wavelength (``bosonized'') form for the density fluctuation
of the molecules $\d\rho_m(r)=\pi^{-1}\partial_x\phi+{\bar\rho}_m\cos(2\pi\rho_m x+2\phi_m(r))$,
plugging it into the effective action (\ref{Sm}). Due to the
spatial derivative, the first term
gives rise to an irrelevant coupling $\propto q^2$. The other component of the density leads to
\be
S_{diss}=-{\tilde g}^2\int dr dr' \rho^2 e^{i q_0 (x-x')} \chi_a(r-r') \cos(2\phi_m(r)-2\phi_m(r'))\label{rhorhod} \ee
where $q_0\equiv 2\pi{\bar\rho}_m$.
 Because of the
fast decay of the kernel, only nearby molecules contribute to this interaction appreciably. Moreover, since we are dealing with dipolar molecules with strong and extended repulsive interactions, the relative positions of nearby molecules are effectively pinned, and we can safely replace the cosine term by
the quadratic expansion of its argument. We obtain the quadratic dissipative
action

\begin{widetext}
\bea
S_{diss}&=& {\tilde g}^2\sum_i\w\int d^2 q \left[\tilde{\chi}_a(q_0,0,0)-\tilde{\chi}_a(q_x+q_0,q_y,\w)\right]\phi_m^\star(q_x,\w)\phi_m(q_x,\w)\nn\\
&=&{\tilde g}^2\sum_{i\w}\int d^2 q \left[{1\over v_a^2}-{ q_y^2+(q_x+q_0)^2\over \w^2+v_a^2q_y^2+v_a^2(q_x+q_0)^2}\right]\phi_m^\star(q_x,\w)\phi_m(q_x,\w)\nn\\
&=&{\tilde g}^2\sum_{i\w}\int {d q_x\over v_a} {\w^2\over \sqrt{\w^2+v_a^2 (q_x+q_0)^2}}\phi_m^\star(q_x,\w)\phi_m(q_x,\w)
\label{Sdiss}
\eea
\end{widetext}
Because the atomic condensate is much more weakly interacting than the dipolar molecules its sound velocity $v_a$ is naturally much smaller than that
of the one dimensional molecular system $v_m$. Therefore for a wide range of frequencies $2\pi{\bar\rho}_m v_m> \w>> 2\pi{\bar\rho}_m v_a$ one can neglect the
$q$ dependence in (\ref{Sdiss}). And so within this frequency range the two dimensional condensate acts as an ohmic bath with
\be
S_{diss}={\tilde g}^2\sum_{i\w}\int d q_x |\w|\phi_m^\star(q_x,\w)\phi_m(q_x,\w).
\ee

Note that immersion in a three dimensional superfluid leads to a different result. The same analysis shows that such a condensate provides "super-ohmic" dissipation with
a frequency dependence $\w^2$. This is not sufficiently strong dissipation to balance the $1/f$ noise, as required to yield the scale invariant steady states.

\section{Derivation of the non equilibrium Response function}

In this section we outline the derivation of equation (5). In real time and space, the retarded density-density linear response function is given by:
\be
\chi(x,t)=i\av{\rho\left(\phi_{f}(x,t)\right)\left[\rho\left(\phi_f(0,0)\right)
-\rho\left(\phi_b(0,0)\right)\right]}_0.
\label{response}
\ee
Here $\av{}_0$ is the expectation value, with respect to the Keldysh action of Eq. (8); $\phi_f(x,t)$ and $\phi_b(x,t)$ are the components of the field on the forward and backward paths.

The nature of the response depends in an important way on the wavelength of the applied perturbation. Long wavelength perturbations (compared to $\rho_0^{-1}$) couple to the smooth, long wavelength component of the density as $V(x)\partial_x\phi$. Since the coupling is linear, the response to it is independent of the response to the noise and therefore the same as in equilibrium. This is just the "superposition principle" in a harmonic system.

A perturbation of wavelength near $q_0$ on the other hand couples non linearly to the harmonic system as $V(x)\cos(2\phi(x))$, via the $q_0$ component of the density fluctuation.
We will see that the resulting response depends on both $\av{\phi_{cl} \hat\phi}$ and $\av{\phi_{cl}\phi_{cl}}$ and it is highly affected by the noise.
Expressing equation (\ref{response}) in terms of classical and quantum components we have:
\be \chi(x,t) = 2\av{\cos(q_0 x + 2\phi_{cl}(x,t))\sin(2\phi_{cl}(0,0))\sin(2\hat\phi(0,0))} \label{eq:chi}\ee

Since the action (8) is quadratic, we can use the identity: $\av{\exp(2i\phi)}=\exp(-2\av{\phi^2})$. Inverting (8) we obtain:
\bea
\av{\phi_{cl}(x,t)\phi_{cl}(0,0)}&=&i\frac{K_*}4 \log((x^2-t^2)/a^2)\nn\\ \av{\phi_{cl}(x,t)\hat\phi(0,0)}&=&-i\frac{\pi K}4 \Theta(t-|x|).
\eea

Here $K_* =  K(1+\pi^{-2} F_0/\eta)$, $\Theta$ is the Heaviside step function and $a$ is a UV cutoff. Equation (\ref{eq:chi}) becomes:
\be \chi(x,t) =\frac14\cos(q_0 x)\left(\frac{a}{x^2-t^2}\right)^{K_*}\sin(\pi K) \Theta(t-|x|)\ee
The response is zero for negative times, as required by causality.

The energy transferred from the probe to the system per unit time is proportional to $\w\chi''(q,\w)$, where $\chi''$ is the imaginary part of the Fourier transform of (\ref{response}). The two dimensional Fourier transform is easily performed by the change of variable $s=(x+t)/\sqrt{2}$ and $t=(x-t)/\sqrt{2}$ and, for positive frequencies, it leads to (5):
\bea
\w \chi''(q,\w)&=&C\left(K,K_\star\right)|\w|(\w^2-\d q^2)^{K_\star-1}\Theta(\w^2-\d q^2)\nn\\
C(K,K_\star)&=&{1\over 4\Gamma^2(K_\star)}{\sin(\pi K)\over\sin(\pi K_\star)}
\label{chipp2}
\eea

\bibliographystyle{naturemag}
\bibliography{noneq}

\end{document}